\documentclass[aps,showpacs,preprint, superscriptaddress]{revtex4}
\usepackage[dvips]{graphicx}
\usepackage{latexsym}
\usepackage{natbib}
\usepackage{amsmath}
\usepackage{color}
\usepackage[percent]{overpic} 

\begin{document}
\title{Cascading failures in interdependent networks\\ with finite
  functional components} 

\author{M. A. Di Muro} \affiliation{Instituto de Investigaciones
  F\'isicas de Mar del Plata (IFIMAR)-Departamento de F\'isica,
  Facultad de Ciencias Exactas y Naturales, Universidad Nacional de
  Mar del Plata-CONICET, Funes 3350, (7600) Mar del Plata, Argentina.}
\author{S. V. Buldyrev} \affiliation{Department of Physics, Yeshiva
  University, 500 West 185th Street, New York, New York 10033, USA}
\author{H. E. Stanley} \affiliation{Center for Polymer Studies, Boston
  University, Boston, Massachusetts 02215, USA}
\author{L. A. Braunstein} \affiliation{Instituto de Investigaciones
  F\'isicas de Mar del Plata (IFIMAR)-Departamento de F\'isica,
  Facultad de Ciencias Exactas y Naturales, Universidad Nacional de
  Mar del Plata-CONICET, Funes 3350, (7600) Mar del Plata, Argentina.}
\affiliation{Center for Polymer Studies, Boston University, Boston,
  Massachusetts 02215, USA}

\begin{abstract}

\noindent We present a cascading failure model of two interdependent
networks in which functional nodes belong to components of size
greater than or equal to $s$. We find theoretically and via simulation
that in complex networks with random dependency links the transition
is first order for $s\geq 3$ and continuous for $s=2$. We also study
interdependent lattices with a distance constraint $r$ in the
dependency links and find that increasing $r$ moves the system from a
regime without a phase transition to one with a second-order
transition. As $r$ continues to increase, the system collapses in a
first-order transition. Each regime is associated with a different
structure of domain formation of functional nodes.

\end{abstract}


\maketitle 

\section{Introduction}

Modern real-world infrastructures can be modeled as a system of several
interdependent networks \cite{Ros_01,Peerenboom,Rinaldi,Yagan}.
For example, a power grid and the communication network that executes
control over its power stations constitute a system of two
interdependent networks. Power stations depend on communication networks
to function, and communication networks cannot function without
electricity. There have been several recent attempts to model these
systems
\cite{Buldyrev2010,Buldyrev2011,Parshani,Shao,Huang,Gao2011,Gao2012,
  Li,Gao2012a,Bashan,Gao2013,Son_01,Bianconi2014,Bax_01,Valdez2013,Valdez2014}.
One of these is based on a model of mutual percolation (MOMP) in which a
node in each network can function only if (1) it receives a crucial
commodity from support nodes in other networks and (2) it belongs to the
giant component (GC) formed by other functional nodes in its own
network.

If the nodes within each network of the system are randomly connected,
and the support links connecting the nodes in different networks are
also random, then the MOMP for an arbitrary network of networks (NON)
can be solved analytically using the framework of generating functions,
which allows to map the stochastic model into node percolation.

It turns out that a NON is significantly more vulnerable than
a single network with the same degree distribution.  
In regular percolation of a single network, the size of the GC gradually
approaches zero when the fraction $p$ of nodes that survived the initial failure,
approaches the critical value $p_c$. In contrast, in the MOMP, the fraction
of nodes in the mutual GC, $\mu(p)$ undergoes a discontinuous first-order phase 
transition at $p=p_\tau>p_c$, dropping from a positive value, $\mu_\tau$, for 
$p\geq p_\tau$ to zero, for $p<p_\tau$.  

The authors of Ref.~\cite{Li} extended MOMP to Euclidian lattices by
studying the process of cascading failures in two lattices $A$ and $B$
of the same size $L$ in which the dependency links are limited by a
distance constraint $r$. In this case there is a particular value of
$r$ denoted by $r_{\rm max}$ below which there is a second-order
transition and above which the system collapses in a first-order
transition. This process is characterized by the formation of spatial
holes that burn the entire system when $r\geq r_{\rm max}$
\cite{note_Mati3}.

The first rule of MOMP is quite general and can be easily verified from
an engineering standpoint, but the second rule is not easy to
verify. Although it seems that a functioning node must belong to the
giant component in order to receive sufficient power, information, or
fuel from its own network, this condition can be relaxed, i.e., the
second rule in the MOMP can be replaced by a more general rule
($2^\prime$) in which a node in order to be functional must belong to a
connected component of size greater than or equal to $s$, formed by other
functional nodes of this network. This rule is significantly more
general and realistic than rule (2) because the nodes in finite
components are still able to receive sufficient commodities to continue
functioning. Note that the original rule (2) is actually a particular
case of rule ($2^\prime$) for $s=\infty$.  In this paper, we will show
how the replacement of condition (2) by the more general condition
$(2^\prime)$ with $s<\infty$ affects the results in complex networks and
Euclidean lattices \cite{Buldyrev2010,Li}. 

\section{Theoretical Formalism for Complex Networks}
The most important role of the MOMP of a NON is 
played by the function $g_i(y_i)$
\cite{Buldyrev2010} such that $y_i g_i(y_i)$ is the fraction of nodes in
the {\it giant component\/} of network $i$ of the NON 
after a random failure of a
fraction $1-y_i$ of its nodes.
The generating function of the degree distribution of network $i$ is
given by \cite{Buldyrev2010,Gao2012}.
\begin{equation}
G_i(x)=\sum_{k=0}^\infty P_{k,i} x^k,
\label{e:1}
\end{equation}   
where $P_{k,i}$ is the degree distribution of network $i$ and the
generating function of the excess degree distribution is
\begin{equation}
H_i(x)=\frac{d}{dx}\frac{G_i(x)}{\langle k_i \rangle} =
\sum_{k=0}^\infty P_{k+1,i}(k+1) x^k/\langle k_i\rangle, 
\label{e:2}
\end{equation} 
where 
\begin{equation}
\langle k_i\rangle=\sum_{k=1}^\infty k\;P_{k,i}=G_i^{'}(x)|_{x=1}
\end{equation}
is the average degree of network $i$. 

The fraction of nodes in the giant component relative to the fraction $y$ of surviving nodes, is given by 
\begin{equation}
g_i(y)=1-G_i[f_i(y)y+1-y],
\label{e:3}
\end{equation}
where $f_i(y)$ is the probability that the branches do not reach the GC, which satisfies the
recursive equation \cite{Bra_01}
\begin{equation}
f_i(y)=H_i[f_i(y)y+1-y].
\label{e:4}
\end{equation}
We also compute the generating function of the component size distribution
\cite{Newman} 
\begin{equation}
C_i(x,y)=\sum_{s=1}^\infty\pi_{i,s}(y) x^s=x G_i[B_i(x,y)y+1-y],
\label{e:5}
\end{equation}
where $\pi_{i,s}(y)$ is the fraction of nodes belonging to components of size $s$
in network $i$ relative to the fraction $y$ of surviving nodes, and
$B_i(x,y)$ satisfies the recursive equation
\begin{equation}
B_i(x,y)=x H_i[B_i(x,y)y+1-y].
\label{e:6}
\end{equation}
Note that when $x=1$, Eqs.~(\ref{e:5}) and (\ref{e:6}) are equivalent
to Eqs.~(\ref{e:3}) and (\ref{e:4}), respectively, and hence
\begin{equation}
C_i(1,y)=\sum_{s=1}^\infty \pi_{i,s}(y)=1-g_i(y).
\label{e:7}
\end{equation}

To move from rule (2) to rule
$(2^\prime)$ we replace function $g_i(y_i)$ with function
$g_{i,s}(y_i)$, defined the same as $g_i(y_i)$ but replacing the words
{\it giant component\/} with {\it components of size larger than or
  equal to $s$}. Thus,

\begin{equation}
g_{i,s}(y)=1-\sum_{r=1}^{s-1} \; \pi_{i,r}(y).
\label{e:8}
\end{equation}

\section{Analytic solution in Random Regular  and  Erd\"{o}s R\'eny  networks}\label{S.anal}

In this section we present the analytic solution for two random
regular (RR) and two Erd\"{o}s R\'eny (ER) interdependent networks.
From Eq.~(\ref{e:8}), using the Lagrange inversion formula ~\cite{Newman}
we obtain the coefficients $\pi_{i,s}(y)$ for $s>1$
\begin{equation}
\pi_{i,s}(y)=\frac{y\langle k_i\rangle
}{(s-1)!}\frac{d^{s-2}}{dx^{s-2}}[H_i(x\;y+1-y)]^s|_{ x=0} \;, 
\label{e:piis}
\end{equation}
and  
\begin{equation}
 \pi_{i,1}(y)=G_i(1-y).
\end{equation}
For ER graphs with a Poisson degree distribution and
an average degree $\langle k \rangle$ and for RR graphs
with degree $z$, we can obtain an analytical solution for
Eq.~(\ref{e:piis}) for $\pi_{i,s}(y)$. For ER networks $\pi_{ER,s}(y)$
is given by
\begin{equation}   
\pi_{ER,s}(y)= \frac {(s\;y\;\langle
    k\rangle)^{s-1}\;\exp(-s\;y\;\langle k\rangle)}{s!}  ,
\label{f:piER}
\end{equation}
and for RR graphs, with degree $z$, for $s=1$,
$\pi_{RR,1}(y)$ is given by
\begin{equation}   
\pi_{RR,1}(y)= (1-y)^z\;,
\end{equation}
and when $s>1$, $\pi_{RR,s}(y)$ is
\begin{equation}   
\pi_{RR,s}(y)=
z\;p^{s-1}(1-y)^{s\;(z-2)+2}\frac{[s\;(z-1)]!}{(s-1)![s\;(z-2)+2]!}.
\label{f:piRR}
\end{equation}

\section{Model in Complex Networks}
To illustrate our model, we consider two networks $A$ and $B$ with
degree distributions in which bidirectional interdependency links
establish a one-to-one correspondence between their nodes as in
Ref.~\cite{Buldyrev2010}.  The initial random failure of a fraction
$1-p$ of nodes in one network at $t=0$ produces a failure cascade in
both networks.  

\subsection{Theory}
At step $t$ of the failure cascade, the effective
fraction of surviving nodes $\tilde{\mu}_{A,t}(p)$ and
$\tilde{\mu}_{B,t}(p)$ of networks $A$ and $B$, respectively,
satisfies the recursive equations
\begin{equation}
   \begin{aligned}
      \tilde{\mu}_{A,t}(p)=p\;g_{B,s} \Big(\tilde{\mu}_{B,t-1}(p) \Big),\\
      \tilde{\mu}_{B, t}(p)=p\;g_{A,s}(\tilde{\mu}_{A, t}(p)),
    \end{aligned}
\end{equation}
and the fractions of nodes belonging to components of size greater than or
equal to $s$, $\mu_{A,t}(p)$ and $\mu_{B,t}(p)$, are given by

\begin{equation}
   \begin{aligned}
      \mu_{A,t}(p)= \tilde{\mu}_{B, t}(p)\;g_{B,s} \Big( \tilde{\mu}_{B, t}(p)\Big),\\
     \mu_{B,t}(p)=\tilde{\mu}_{A, t}(p)\;g_{A,s}(\tilde{\mu}_{A, t}(p)),
    \end{aligned}
\label{e:mu00}
\end{equation}
 where $\tilde{\mu}_{A,0}(p)=p$ and $\mu_{A,0}(p)=p\;g_{A,s}(p)$.  The
  process is iterated until the steady state is reached, where
\begin{equation}
   \begin{aligned}
      \tilde{\mu}_{A}(p)=p\;g_{B,s}(\tilde{\mu}_{B}(p)),\\
      \tilde{\mu}_{B}(p)=p\;g_{A,s}(\tilde{\mu}_{A}(p)),
    \end{aligned}
\label{e:mu01}
\end{equation}
and
\[
\mu\;(p)\equiv\mu_{A}(p)=\mu_{B}(p)=\tilde{\mu}_{A}(p)\tilde{\mu}_{B}(p)/p.
\]
When $p=p_\tau$, the order parameter of our model, $\mu(p)$, transitions
from $\mu(p)>0$ when $p>p_\tau$ to $\mu(p)=0$ when $p \leq p_\tau$. In the
most simple case when the networks have identical degree
distributions, $g_{A,s}(x)=g_{B,s}(x) \equiv g_{s}(x)$. At the
threshold, $p=p_\tau$ and $\tilde{\mu}(p_\tau)$ satisfy
\begin{equation}
\begin{aligned}
\tilde{\mu}(p_\tau)&=&p_\tau\;g_{s}\left( \tilde{\mu}(p_\tau) \right)\\ 1
&=&p_\tau\;g_{s}^\prime(\tilde{\mu}(p_\tau)),
\end{aligned}
\label{e:mu1}
\end{equation}
where $g_{s}^\prime(y)=d g_{s}(y) /dy$. Because $g_2(y)=1-G(1-y)$, the
second derivative of $g_2(y)$ is always negative, and thus
Eq.~(\ref{e:mu1}) has a trivial solution at $\tilde{\mu}(p_\tau)=0$
from which $p_\tau=1/G^\prime(1)=1/\langle k\rangle$, where
$G^\prime(1)= d G(y)/d y|_{y=1}$, and as a consequence the system
undergoes a continuous phase transition. For networks with a
non-divergent second moment of the degree distribution the transition
is third-order, but for networks with a divergent second moment
the transition is of a higher order. However, when $s \geq 3$,
$g_s(y)$ changes the sign of its second derivative from positive at
$y=0$ to negative at $y=1$, and hence Eq.~(\ref{e:mu1}) has a
nontrivial solution in the interval $0<p<1$ at which $\tilde{\mu}(p)$
abruptly changes from a positive value above $p_\tau$ to zero below
$p_\tau$.  Thus for $s \geq 3$ we always have a first-order
transition, which was previously found \cite{Buldyrev2010}, but only
for $s=\infty$.  The different kinds of transitions that we find in
our model are reminiscent of the ones found in {\it k}-core percolation
\cite{Doro_2002,Doro_2008,Bax_2010,Bax_2011}.  A {\it k}-core of a graph is
a maximal connected subgraph of the original graph in which all
vertices have degree at least $k$, formed by repeatedly deleting all
vertices of degree less than $k$. In particular, in 2-core there is a
continuous transition, while for $k\geq 3$ the transition is
first-order, as in our model for $s=2$ and $s\geq 3$ respectively.
The key difference between the {\it k}-core transition and our model is that
in our model the functionality of a node is not based on its degree
but rather on the size of the finite components to which it
belongs. The similarity between the phase transitions in our model and
the ones in {\it k}-core is due to a resemblance between the pruning rules
of both processes. For example, in our model with $s=2$, the final
state is constituted of nodes with at least one active link in their
own network and one dependency link and, hence, all nodes have two active
links as in the final state of 2-core. Next we will see that the
similarities of the phase transitions arise due to the similarities in
the leading terms of the Taylor expansions of the equations that
govern {\it k}-core and our model. However, we will also demonstrate that
both models do not belong to the same universality class.


\subsubsection{Scaling behavior of the fraction of active nodes for $s=2$ in our model}
From Eq.~(\ref{e:mu1}) for $s= 2$, at the steady state, the effective
fraction of remaining nodes $\tilde{\mu}(p) \equiv \tilde{\mu}$ is
given by
\begin{equation}\label{Eq.0}
\tilde{\mu}=p[1-G(1-\tilde{\mu})],
\end{equation}
where $p$ is the fraction of nodes that survived the initial damage,
and $G(x)$ is the generating function of the degree distribution.
For RR, ER, and scale-free networks with nondivergent second moment
($\lambda> 3$), close to the threshold $p_\tau$ at which $\tilde{\mu}(p_\tau)=0$,
expanding Eq.~(\ref{Eq.0}) around $\tilde{\mu}=0$ gives

\begin{equation}\label{Eq.2}
\tilde{\mu}= p[G^\prime(1) \tilde{\mu}- G^{\prime\prime}(1) \tilde{\mu}^2/2 +O(\tilde{\mu}^{3})],
\end{equation}

and solving this equation for $\tilde{\mu}$ leads to

\begin{equation}\label{Eq.3q}
\tilde{\mu} = 2 \frac{p G^\prime(1)-1}{ p\; G^{\prime\prime}(1)}
+O(\tilde{\mu}^{2}).
\end{equation}
Equation~(\ref{Eq.3q}) shows that $\tilde{\mu}\to 0$, when $p\to
1/G^\prime(1)$; thus there is a continuous phase transition at
$p=p_\tau\equiv 1/G^\prime(1)$.  Recalling that for any degree
distribution with converging first and second moments,
$G^\prime(1)=\langle k \rangle$, $G^{\prime\prime}(1)=\langle k^2
\rangle - \langle k \rangle$, we can rewrite Eq.~(\ref{Eq.3q}) as

\begin{equation}
\tilde{\mu} = 2 \frac{\delta p \langle k \rangle}{(p_\tau+\delta p)(\langle
  k^2 \rangle - \langle k \rangle)} +O((\delta p)^2),
\end{equation}
where $p=p_\tau+\delta p$, with $\delta p \to 0$. Since the denominator
does not diverge then $\tilde{\mu} \sim (p-p_\tau)^{\beta^{'}}$, with
$\beta^{'}=1$.

For two interdependent networks with the same degree distribution, the
order parameter is given by

\begin{equation}\label{Eq.mu}
\mu=\tilde{\mu}^2/p,
\end{equation}

and thus $\mu \sim (p-p_\tau)^\beta$ with $\beta=2$.

For $2<\lambda<3 $, the second moment diverges, thus using the
Tauberian theorem~\cite{Feller_1972} the expansion of $\tilde{\mu}$ is
given by
\begin{equation}
\tilde{\mu}= p[G^\prime(1) \tilde{\mu} - A \tilde{\mu}^{\lambda -1} +O(\tilde{\mu}^{\lambda-2})],
\end{equation}
in which, for $p=p_\tau+ \delta p$,
\begin{eqnarray}
\tilde{\mu}&=&\left(\frac{ \delta p G^\prime(1)}{pA}\right)^{1/(\lambda-2)}+O(\delta p)^{1/(\lambda-2)}\\
      & \sim& (p-p_\tau)^{1/(\lambda-2)},
\end{eqnarray}
so $\beta^{'}=1/(\lambda-2)$, and as a consequence [See
  Eq.~(\ref{Eq.mu})] $\beta=2/(\lambda-2)$. Thus there is a
fourth-order phase transition for $5/2<\lambda<3$.
In general, for scale-free (SF) networks with $2 <\lambda < 3$, the transition is of
$m$th order for $2+2/m<\lambda<2+2/(m-1)$.

\subsubsection{Scaling behavior of the fraction of active nodes in 2-core}

In contrast with Eq.~(\ref{Eq.0}), for 2-core percolation, the fraction
of active nodes $q$ obeys the equation
\begin{equation}\label{Eq.3}
q=p[1-G(1-f)-fG^\prime(1-f)],
\end{equation}
where $p$ is the fraction of nodes that survived the initial damage
and $f$ is the effective fraction of survived links obeying a
self-consistent equation
\begin{equation}\label{Eq.4}
f=p\left[1-\frac{G^\prime(1-f)}{G^\prime(1)}\right].
\end{equation}
For homogeneous networks, such as RR and ER, after expanding
   Eq.~(\ref{Eq.4}) around $f=0$, we obtain
\begin{equation}
f=\frac{p}{G^\prime(1)}[G^{\prime\prime}(1)f-G^{\prime\prime\prime}(1)f^2/2 +O(f^3)].
\end{equation}
If $G^{\prime\prime\prime}(1) < \infty$, then $p_\tau=G^\prime(1)/G^{\prime\prime}(1)= \langle
k\rangle/( \langle k^2\rangle-\langle k\rangle)$ as in regular
percolation, and
\begin{equation}
f=\frac{2[\delta p G^{\prime\prime}(1)]}{p\;G^{\prime\prime\prime}(1)}.
\end{equation} 

Finally expanding Eq.~(\ref{Eq.3}) around $f=0$ leads to
$q=pf^2G^{\prime\prime}(1)/2+O(f^3)\sim (p-p_\tau)^2$, which indicates a
third-order phase transition.

For SF networks, if $3<\lambda \leq 4$, from the Tauberian
theorem~\cite{Feller_1972}
\begin{equation}
f=p\left[\frac{G^{\prime\prime}(1)}{G^{\prime}(1)}f-Af^{\lambda-2}+O(f^{\lambda-2})\right],
\end{equation}
from where $f\sim (p-p_\tau)^{1/(\lambda -3)}$ with
$p_\tau=G^{\prime}(1)/G^{\prime\prime}(1)$ and
$q\sim(p-p_\tau)^{2/(\lambda-3)}$ and the transition becomes of the
order $m$ if $3+2/m<\lambda \leq 3+2/(m-1)$. If $2 < \lambda < 3$,
$G^{\prime\prime}(1)=\infty$, then $p_\tau =0$, $f\sim
p^{1/(3-\lambda)}$ and $q\sim p^{1/(3-\lambda)}$.  Thus for
$2<\lambda<3$ there is a phase transition but at $p_\tau=0$, and the
order parameter of this transition changes in reverse order from
infinity for $\lambda=3$ to $3$ for $\lambda =2+\epsilon$ with
$\epsilon \to 0$. 

Thus we have a close analogy between the model of functional finite
component interdependent networks with $s=2$ and $2$-core percolation
in terms of the order of the phase transition. This analogy stems from
the similarities in the Taylor expansion of the equations describing
these two models, but the physical basis on which these equations are
constructed totally differs. In addition, the order of the
transitions differs for SF networks with $2<\lambda<4$, and thus
the two models do not belong to the same universality class.

\subsection{Simulations in complex networks}
We test our theoretical arguments with stochastic simulations in which
we use the Molloy-Reed algorithm \cite{Mol_01} to construct networks
with a given degree distribution. The procedure is as follows:

(1) At $t=0$ we remove a random fraction of nodes $1-p$ in
  network $A$, remove all the nodes in the components of network $A$
  smaller than $s$, and remove all the dependent nodes in network $B$.
 
(2) At $t \geq 1$ we remove all the nodes in the components of
  network $B$ smaller than $s$ and remove all the nodes in network $A$
  dependent on dead nodes in $B$.

(3) We repeat (2) until no more nodes can be removed.

\begin{figure}[h]
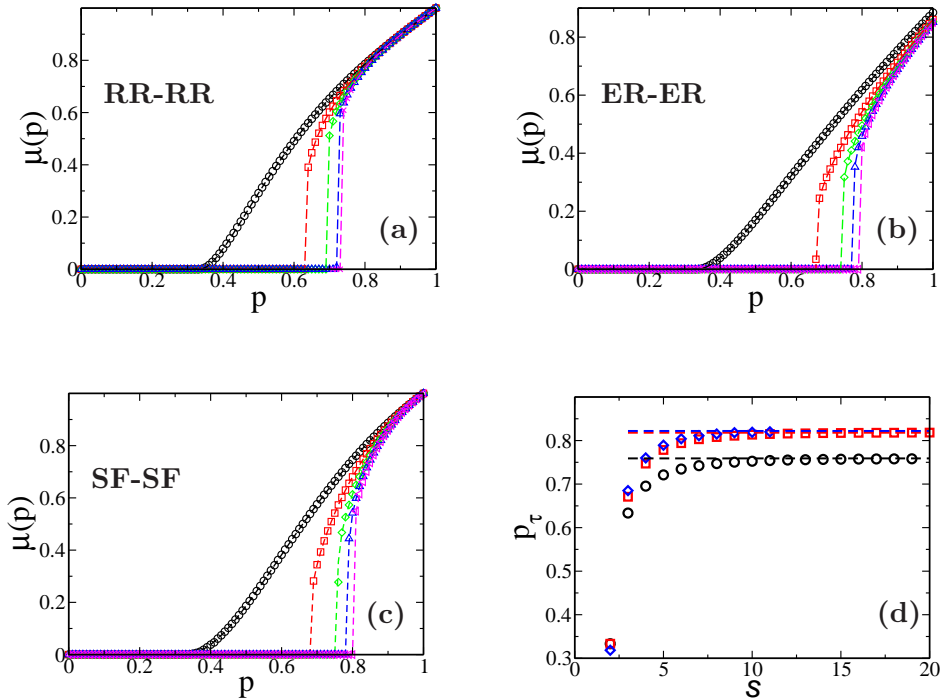
 
\vspace{1cm}
\begin{center}
  \begin{overpic}[scale=0.23]{Fig1a.eps}
    \put(85,18){\bf{(a)}}
\put(20,50){\bf{RR-RR}}
\end{overpic}\hspace{1cm}\vspace{1cm}
  \begin{overpic}[scale=0.23]{Fig1b.eps}
    \put(85,18){{\bf{(b)}}}
\put(20,50){{\bf{ER-ER}}}
  \end{overpic}\hspace{1cm}\vspace{1cm}
 \begin{overpic}[scale=0.23]{Fig1c.eps}
    \put(85,18){{\bf{(c)}}}
\put(20,50){{\bf{SF-SF}}}
  \end{overpic}\hspace{1cm}
 \begin{overpic}[scale=0.23]{Fig1d.eps}
    \put(85,18){{\bf{(d)}}}
  \end{overpic}
\bigskip
\end{center}
\vspace{-2cm}
\caption{$\mu(p)$ as a function of $p$ for different values of $s$ for
  two (a) RR networks with degree $z=3$ for $N=10^6$ , (b) ER networks
  with average degree $\langle k \rangle=3$ for $N=10^6$, and (c) SF
  networks with $\lambda=3$ for $N=5\;10^6$ with $1000$ networks
  realizations for different values of $s$, from $s=2$ to $s=6$, (the
  left most curves, indicated by circles). The symbols are the
  simulations and the dashed lines are the theory. (d) The threshold $p_\tau$
  as a function of $s$ obtained from the theory for the same RR
  (black, circles), ER (red squares), and SF (blue diamonds) networks
  presented in (a), (b), and (c). The dashed lines are used as a guide
  to show $p_\tau$ for $s \to \infty$.}\label{Fig.2}
\end{figure}

We perform simulations for a system of two ER graphs, two RR graphs in
which all nodes have the same degree $z$, each of $N=10^6$ nodes, and
two SF graphs with $N= 5 \times 10^6$ (see
Fig.~\ref{Fig.2}). The SF networks have a degree distribution $P_k
\propto k^{-\lambda}$ with $k_{\rm min}\leq k \leq k_{\rm max}$, where
$\lambda$ is the exponent of the SF network. We set $k_{\rm min}=2$
and $k_{\max}= \sqrt{N}$. To compare our simulations with the
theoretical results [Eq.~(\ref{e:mu1})] we use analytical expressions
for $\pi_{i,s}(p)$ given in the case of ER and RR networks by
Eq.~(\ref{e:piis}).  For SF networks we compute $\pi_s(p)$
numerically. The details of the analytical solution for ER and RR
networks are presented in Sec.~\ref{S.anal}.

Figures \ref{Fig.2}(a), \ref{Fig.2}(b), and \ref{Fig.2}(c) show perfect
agreement between the theoretical results and the simulations.
Figure~\ref{Fig.2}(d) shows a plot of $p_\tau$ as a function of $s$ for
two RR networks with degree $z=3$, two ER networks with $\langle k
\rangle=3$, and two SF networks with $\lambda=3$, $k_{\rm min}=2$ and an
average degree $\langle k \rangle=3.18$. As predicted, $p_\tau=1/\langle
k \rangle$ for $s=2$ and increases as $s$ increases.  For $s \to \infty$
we recover the mutual percolation threshold of Ref.~\cite{Buldyrev2010}
shown as dashed lines in Fig.~\ref{Fig.2}(d).  

\section{Model in Interdependent Euclidean lattices}
We also study the same model for square lattices, generalizing
Refs.~\cite{Li,Bashan}. 
When there are random interdependency links,
i.e., when there is no geometric constraint on the interdependencies,
we use the exact results for the perimeter polynomials of the finite
components to compute $g_s(p)$,
\begin{equation}
g_s(p)= 1-\sum_{n=1}^{s-1}\;n\;p^{n-1}\;D_n(1-p),
\end{equation}
where $D_n(1-p)$, are the perimeter polynomials for small components
on a square lattice \cite{Sykes}.

\begin{figure}[h]
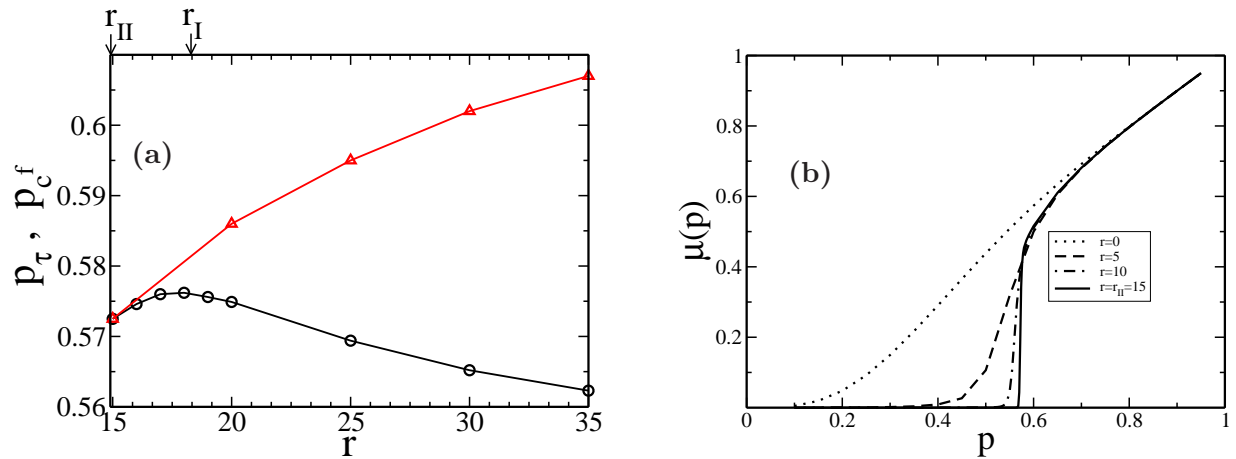
 
\vspace{1cm}
\begin{center}
  \begin{overpic}[scale=0.31]{Fig2a.eps}
    \put(20,50){\bf{(a)}}
\end{overpic}\hspace{1cm}\vspace{1cm}
  \begin{overpic}[scale=0.31]{Fig2b.eps}
    \put(20,50){{\bf{(b)}}}
  \end{overpic}
\vspace{1cm}
\end{center}
\vspace{-3cm}
\caption{For interdependent lattices with interdependent distance $r$
  and survival component size $s=4$, for $L=512$ (a) $p_\tau(r)$
  ($\bigcirc$) vs. $r$ for the first-order $r \geq r_{I}=18$ and the
  continuous phase transition $r_{II}=15 \leq r < r_I=18$ and $p^f_c(r)$
  ($\bigtriangleup$). The lines are used as a guide for the
  eyes. (b) $\mu(p)$ vs. $p$ for different values of $r$.  For $r<
  r_{II}=15$ the results do not depend on the lattice size $L$. The
  system size dependence emerges only at $r=r_{II}=15$.}
\label{f:Pc4}
\end{figure}

Here the system undergoes a first-order phase transition when $s\geq
3$ at the predicted values of $p_\tau=0.485$ for $s=3$ and $p_\tau=0.5506$
for $s=4$, obtained by solving Eq.~(\ref{e:mu1}). When the
interdependency links satisfy distance restrictions, we define the
distance between the two interdependent nodes in lattices $A$ and $B$
as the shortest path between the nodes along the bonds of the
lattices, i.e., $|x_A-x_B|+|y_A-y_B|\leq r$, where $(x_A,y_A)$ and
$(x_B,y_B)$ are the coordinates of the interdependent nodes in
lattices $A$ and $B$, respectively. Using simulations we see a
first-order phase transition emerging at a certain value of $r=r_{I}$
in qualitative agreement with the case $s=\infty$ studied by Li {\it et
al}.~\cite{Li}. At this value of $r$ the system reaches maximum
vulnerability, indicated by a maximum of $p_\tau(r)$ as a function of $r$
[see Fig.~\ref{f:Pc4}(a)].

The $r_I$ value is much greater than the value obtained for the MOMP
$(s=\infty)$. For $r$ close to $r_I$, the cascading failures propagate
via node destruction on the domain perimeters composed of surviving
node components, and this creates moving interfaces when the size of
the void separating the domains is greater than $r$. These moving
interfaces belong to the class of depinning transitions characterized
by a threshold $p=p_c^f(r)$ that increases with $r$ (see
Fig.~\ref{f:Pc4}).  Here $p=p_c^f(r)$ is the critical fraction of
nodes remaining after the initial failure, such that for $p>p_c^f(r)$
the interface of an infinitely large void will be eventually pinned
and stop to propagate.  In contrast, when $p< p_c^f$, the interface of
the voids propagates freely without pinning and eventually burns the
entire system.  Near $p_c^f(r)$, the velocity of the domain interfaces
approaches zero with a power-law behavior $v\sim (p_c^f-p)^\theta$,
where $\theta>0$ is a critical exponent \cite{Barabasi1995}.  In order
to compute $p_c^f$, we compute the velocity $v$ of the growing
interface as a function $p^f-p$ until we get a straight line in a
log-log plot, which corresponds to the value of the critical threshold
$p_c^f$. The value of the slope of $v \sim p_c^f-p$ is the critical
exponent $\theta$. We find $\theta=0.53$, suggesting that the
interface belongs to the universality class of a Kardar-Parisi-Zhang
(KPZ) equation \cite{KPZ} with quenched noise.  As $p=p_c^f(r)$
increases, the probability that large voids with a diameter greater
than $r$ will spontaneously form, decreases, and becomes vanishingly
small in a system of a finite size. Thus in a finite system we must
decrease $p$ below $p_c^f(r)$ in order to create these voids.  When
$p< p_c^f(r)$, the interface of the voids begins to freely propagate
without pinning and eventually, like a forest fire, burns through the
entire system. Thus the emergence of a first-order transition in a
finite system depends on the system size, i.e., the larger the system,
the larger the $r_I$ value at which the effective first-order
(all-or-nothing) transition is observable.

\begin{figure}[h] 
\vspace{1cm}
\begin{center}
  \begin{overpic}[scale=0.15]{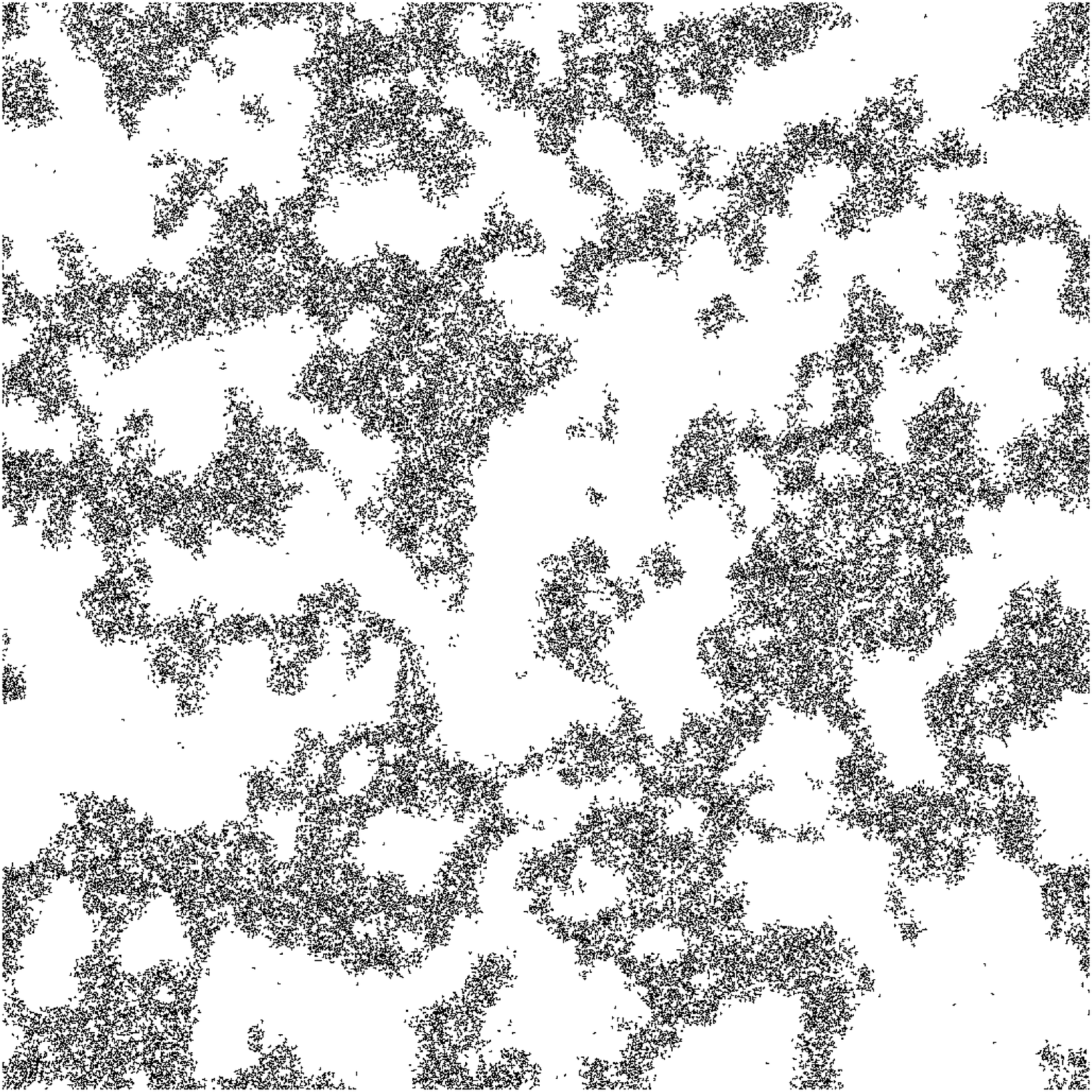}
    \put(20,50){\bf{}}
\end{overpic}\hspace{1cm}
  \begin{overpic}[scale=0.15]{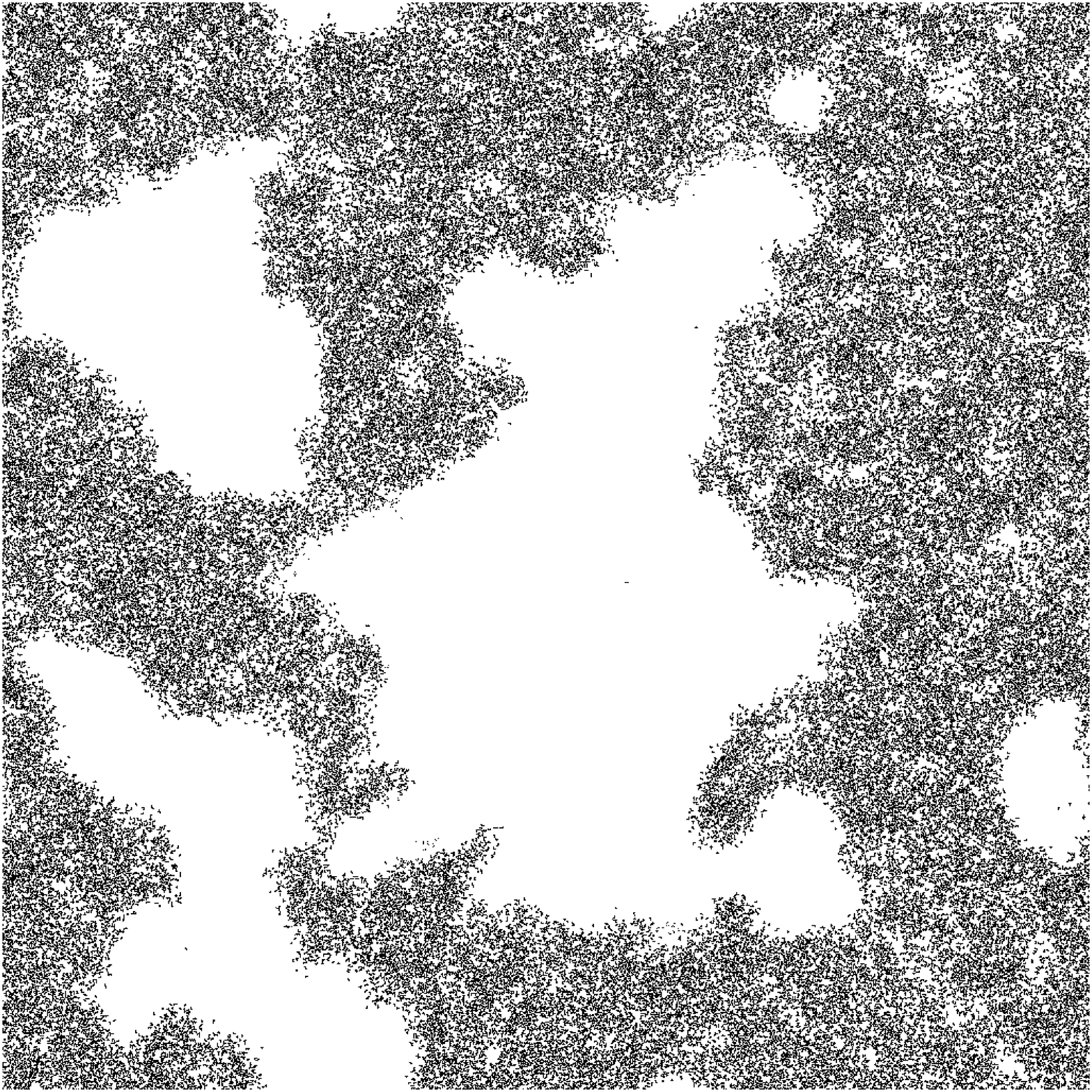}
    \put(20,50){{\bf{}}}
  \end{overpic}
\end{center}
\caption{Snapshots of the model of interdependent lattices for
  $s=4$, $L=1024$ and different values of $r=10 < r_{II}$, $p=0.56$
  (left) and $r=r_{II}=15$, $p=p_m=0.572$ (right) at the end of the
  cascade of failures. It can be seen that for small $r<r_{II}$ the
  system is divided into many independent domains, while for $r=r_{II}$
  the domains coalesce, and the cascades are driven by the propagation
  of the interface near the depinning transition.}
\label{f:picture}
\end{figure}

\begin{figure}[h]
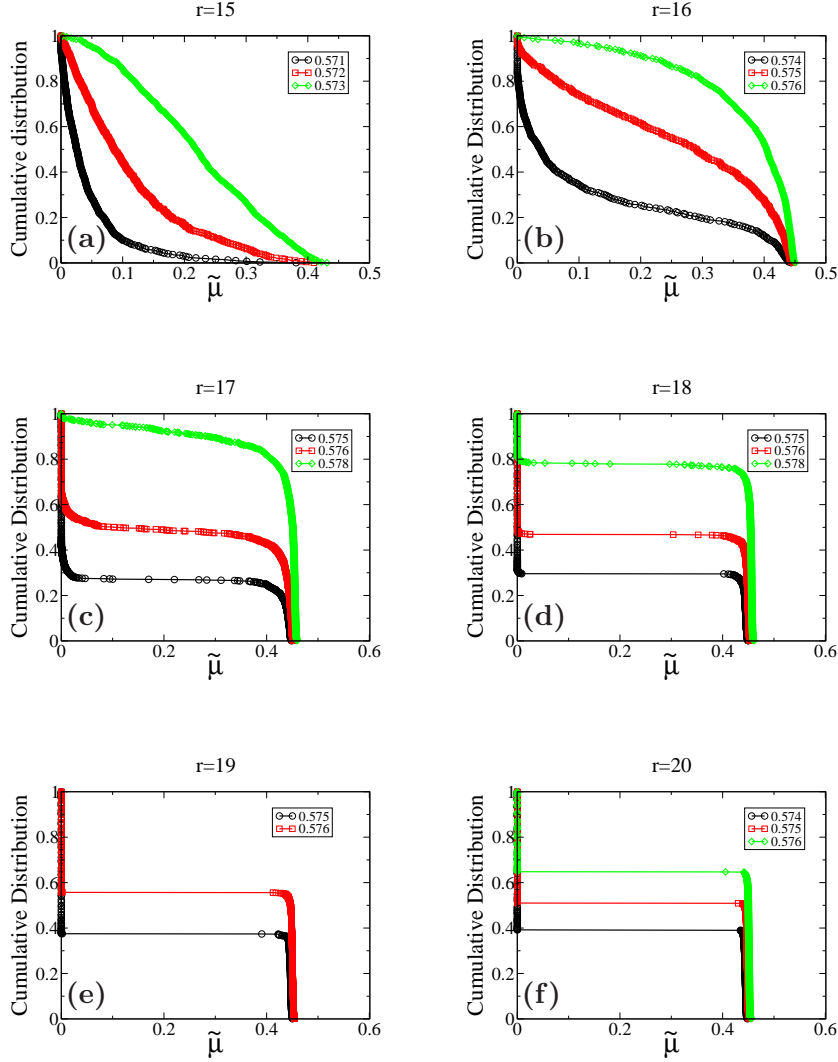
 
\begin{center}
  \begin{overpic}[scale=0.20]{Fig4a.eps}
    \put(15,15){\bf{(a)}}
\end{overpic}\hspace{1cm}
  \begin{overpic}[scale=0.20]{Fig4b.eps}
    \put(15,15){{\bf{(b)}}}
  \end{overpic}\vspace{1cm}\\
  \begin{overpic}[scale=0.20]{Fig4c.eps}
    \put(15,15){{\bf{(c)}}}
  \end{overpic}\hspace{1cm}
  \begin{overpic}[scale=0.20]{Fig4d.eps}
    \put(15,15){{\bf{(d)}}}
  \end{overpic}\vspace{1cm} \\
  \begin{overpic}[scale=0.20]{Fig4e.eps}
    \put(15,15){{\bf{(e)}}}
  \end{overpic}\hspace{1cm}
  \begin{overpic}[scale=0.20]{Fig4f.eps}
    \put(15,15){{\bf{(f)}}}
  \end{overpic}\vspace{1cm}
 \end{center}
\vspace{-1cm}
\caption{Cumulative distribution of the fraction of survived nodes,
  $\tilde{\mu}$ , for different values of $r$. As we can see from the
  plots, as $r$ increases above $r_{II}$, a plateau develops in the
  cumulative distribution for $p \approx p_\tau$, which means that the
  distribution of the values of $\tilde{\mu}$ is bimodal and the
  system will eventually reach a first-order transition at $r \geq
  r_I$. In this regime, there is a large gap between the values of
  $\tilde{\mu}$, indicating that for the same value of $p$, either a
  large fraction of the system can stay functional or the system can
  completely collapse.}
\label{f:Pmu}
\end{figure}

Figure~\ref{f:Pc4}(a) shows that as $r$ continues to increase,
$p_\tau$ begins to decrease and slowly approaches the $p_\tau$ value
for random interdependence as $r\to\infty$.  There is no second-order
percolation transition for finite $s$ and small $r$ that governs the
size of the voids, in contrast to what was found by Li {\it et
  al.}~\cite{Li} for $s\to\infty$.  For finite $s$, a second-order
transition emerges when the $r$ value is large, $r=r_{II}<r_{I}$, but
when $r < r_{II}$ there is no transition, the fraction of survived
nodes $\mu(p)$ is zero only at $p=0$, and it continues to be
differentiable and independent of the system size for any positive
value of $p$. Note, however, that as $r$ approaches $r_{II}$ the
derivative of $\mu(p)$ develops a sharp peak at a certain value of
$p\equiv p_m(r)$ below which $\mu(p)$ is very small but finite. At
$r=r_{II}$ we see a second-order transition because the height of the
peak of the derivative of $\mu(p)$ now increases with the lattice size
$L$, which is typical of a second-order transition.  This behavior is
associated with different regimes of domain formation. For small
values, $r<r_{II}$, the first stages of the cascading failure fragment
the system into small independent regions, each of which has its own
pinned interface(see Fig.~\ref{f:picture}). In this regime, after the
first stages of the cascade of failures the system practically does
not change. After the first stages, the interfaces propagate very slow
and can stop at any point leaving the resulting snapshots
indistinguishable from the one obtained in the steady state. A single
interface emerges only when these regions coalesce at $r=r_{II}$, and
a second-order phase transition related to the propagation of this
interface through the entire system emerges. This second-order phase
transition observed for $r=r_{II}$ has a unimodal distribution of the fraction of surviving nodes $\tilde{\mu}(p)$, and we use the maximum slope of the graph
$\tilde{\mu}(p)$ to compute the critical point $p_\tau=p_m(r_{II})$.  As $r$
increases between $r_{II}$ and $r_{I}$, the distribution of $\tilde{\mu}(p)$ becomes bimodal, and we compute the transition point $p_\tau$ using
the condition of equal probability of both modes.  Note that $p_\tau$
reaches a maximum at $r=r_{I}$ where the two peaks of the distribution
of $\tilde{\mu}(p)$ separate completely, as indicated by a wide plateau in the
cumulative distribution of $\tilde{\mu}(p)$ \cite{Kornbluth}.  The cumulative
distribution of $\tilde{\mu}(p) \equiv \tilde{\mu}$ for square
lattices is presented in Fig.~\ref{f:Pmu}

The emergence of the first order-phase transition above $r_{II}$ is
related to the decrease of the correlation length as we move away from
$r_{II}$. We thus find that when $s$ is small, $r_I$ is
significantly larger than $r_I(\infty)$. For the shortest path metric
$r_I(\infty) =11$, and $r_{II}(4)=15$ and $r_I(4)=18$ for $L=1024$. As
$s$ increases, $r_I$ gradually decreases and coincides with
$r_I(\infty)$ for $s\to\infty$.

\section{Conclusion}
In summary, we find that in complex networks with $s>2$, our model has
a first order transition as for the previously studied case of MOMP
with $s\to \infty$. For $s=2$, our model has a
higher-than-second-order transition similar to that found in {\it
  k}-core, but the order of the transitions in SF networks differs
depending on the exponent of the degree distribution.  However, the
finite component generalization of MOMP in spatially embedded networks
has a totally different behavior, which is not related to {\it
  k}-core. In this case, the transitions, when they exist, are
dominated by the behavior of the pinning transition of void's
interfaces. Our model in spatially embedded networks is a rich and
interesting phenomenon, which has many practical applications for
studying the cascade of failures in real-world infrastructures
embedded in space.  Our work can be extended to any NON model
incorporating MOMP, but our finite component model is significantly
more general and realistic.  We can generalize our model to derive
equations for a partially interdependent NON. Here the second-order
transition will also appear when $s>2$ if the fraction of
interdependent nodes is small.  The value of $s$ can differ in
different networks of the NON and can be a stochastic variable, such
that a component of size $s$ survives with probability $p(s)$, as in
the heterogeneous {\it k}-core~\cite{Bax_2011,Cellay_13}.

\acknowledgments The Boston and Yeshiva University work was supported by DTRA
Grant No.  HDTRA1-14-1-0017, by DOE Contract No. DE-AC07-05Id14517;
and by NSF Grants No. CMMI-1125290, No. PHY-1505000, and
No. CHE-1213217.  S.V.B acknowledges the partial support of this
research through the B.W. Gamson Computational Science Center at
Yeshiva University. L.A.B and M.A.D.M. thank UNMdP and FONCyT, Pict
0429/13, for financial support.


\begin{thebibliography}{35}
\bibitem[{not()}]{note_Mati3}
\bibinfo{note}{Note that the case of random interdependence is the special case
  of $r=L$.}
\end{thebibliography}

\end{document}